\documentclass[12pt]{article}
\usepackage{psfig}
\usepackage{epsf}

\begin{document}

\title{ {\bf Molecular dynamics study of melting of a bcc metal-vanadium I :
mechanical melting}  }

\author{ V.Sorkin, E. Polturak and Joan Adler }
\date{ Dept. of Physics,~Technion Institute of Technology,~32000 Haifa,~Israel}
\maketitle
\begin{abstract}
We present molecular dynamics simulations of the homogeneous
(mechanical) melting transition of a bcc metal, vanadium. We study
both the nominally perfect crystal as well as one that includes
point defects. According to the Born criterion, a solid cannot be
expanded above a critical volume, at which a 'rigidity catastrophe'
occurs. This catastrophe is caused by the vanishing of the elastic
shear modulus. We found that this critical volume is independent
of the route by which it is reached whether by heating the crystal,
or by adding interstitials at a constant temperature which expand the
lattice. Overall, these results are similar to what was found
previously for an fcc metal, copper.  The simulations establish a
phase diagram of the mechanical melting temperature as a function
of the concentration of interstitials. Our results show that the
Born model of melting applies to bcc metals in both the nominally
perfect state and in the case where point defects are present.
\end{abstract}


\section{\label{sec:level1}  Introduction}

 Over the years, several theories explaining the mechanism of melting have been proposed.
 \cite{Ubbelohde,Lind,Ida} This research has by now evolved to a state where
 a clear distinction exists between two possible scenarios for the melting
 transition: a first scenario of {\it homogeneous, or mechanical melting} resulting from
 lattice instability~\cite{ Born, Tallon, Wolf}
 and/or a spontaneous generation of thermal defects,\cite{
Granato,Weber,Cahn1,Cahn2, Kanigel}
 and a second scenario of {\it heterogeneous, or thermodynamic melting} which begins at
 extrinsic defects such as a free surface or an internal
 interface (grain boundaries, voids, etc).\cite{Frenken,Veen,Traynov,Barnett,
 Chen} Throughout this paper we will use the term mechanical melting to describe
 the former case, which we consider here.
  In particular, we take the view proposed by Born
 that at the melting point a {\it 'rigidity catastrophe'}
 is caused by the vanishing of one of the  elastic shear moduli,\cite{Born, Tallon}  $C_{44}$,
 or $C' = (C_{11}-C_{12})/2$.
 In other words, the crystal melts once
 it loses its ability to resist shear.
 This condition determines the mechanical melting temperature, $T_s$, of
 a perfectly homogeneous bulk crystal and was confirmed in extensive studies
 of fcc metals.\cite{Wang1, Wang2, Jin, Kanigel, Cahn2}

   Tallon\cite{Tallon} pointed out that a
 mechanical instability arises when the solid expands up to {\it a
critical specific volume}
 which is close to that of the liquid phase (melt).
 In the study by Wang et al.\cite{Wang1, Wang2} of
 the mechanical melting transition of an fcc solid
under external stress, it was found that volume expansion is
the underlying cause of lattice instability. Kanigel et
al.~\cite{Kanigel} confirmed this scenario in a simulation of fcc
copper in the presence of point defects. They showed that the
critical volume at which a crystal of copper melts is independent
of the path through phase space by which it is reached, whether by
heating of the perfect crystal or by adding point defects
to expand the solid at a constant temperature~\cite{Kanigel}.
 
Solids can undergo mechanical melting only if they have no extended
defects~\cite{Wolf},
 a situation which is conveniently realized in three-dimensional
 computer simulations by means
 of periodic boundary conditions which eliminate the surface.
 Simulations of atomic dynamics for solids with an
 fcc \cite{Wang1, Wang2, Jin, Kanigel, Cahn2} or
 diamond \cite{Phill} structures
 show, among other things, the onset of a shear instability of
 the solid at a temperature $T_s$, which can exceed the thermodynamic melting
temperature $T_m$ by some 20\%, depending on the details of the potential.

 Given the considerable degree of understanding of the melting process
of fcc crystals, it is of substantial interest to see if the scenario of mechanical melting also applies to 
solids having a different lattice structure. We therefore decided to study mechanical
melting of a bcc metal, vanadium, by means of computer simulations.
 We present details of the calculations in Sec. II.
 In Sec. III we describe the results of the simulation of some physical properties
 of a vanadium with and without point  defects. In Sec. IV we present molecular dynamics
 simulation results for mechanical melting in the presence of point
defects.
 Finally, in Sec. IV, we discuss the implication of our results for the
question of the microscopic mechanism of melting.

\section{\label{sec:level1}  Simulation details  }
We model the bulk melting transition of vanadium using the
molecular dynamics simulation~\cite{Rapaport} technique. The
choice of vanadium has no special significance as we are only
interested in the generic features of metallic solids with a bcc
lattice symmetry. While various many-body potentials for fcc
metals ~\cite{Daw} have been developed and thoroughly
tested in numerous simulations, the situation with such potentials
for bcc metals is not as good. This can be explained by the more
complicated nature of the bcc metals in comparison with fcc ones
which  manifests itself in the wide range of elastic constants
(and even in the negative values of the Cauchy condition
$C_{12}-C_{44} <0 $ for some bcc metals). The packing density of
atoms in a bcc lattice is smaller than in a fcc lattice (there are
8 nearest-neighbors in a  bcc lattice and 12 nearest-neighbors in
a fcc). However, the second nearest-neighbor distance  in the bcc
structure is larger than the first nearest-neighbor distance by
only about 15\%. Therefore, the interaction between the second
order and the first order nearest-neighbors in bcc metals is not
negligible, even with screening.
 
In addition, band structure effects are crucial for bcc metals.
A simple approximation which assumes that the electron density can be considered as a superposition
of atomic orbitals is successful for fcc metals, but less appropriate for bcc metals.
Therefore, for metals with the bcc structure electron density is chosen to be an adjustable function,
rather than a superposition of atomic orbitals. Furthermore, angle dependent interactions
could be very important in bcc solids.

For our simulations, we chose the many-body interaction potential developed
by Finnis and Sinclair~\cite{Finnis} ({FS}),
and modified by Rebonato et al.~\cite{Rebonato}
FS proposed a way to incorporate
the delocalized physical nature of the metallic bonding and the essential
band character of bcc metals in a simple model.
The FS potential involves two short ranged
potentials, a cohesive one and a repulsive one. The cohesive potential is
summed over
neighbors and the square root of the result describes the bonding energy.
The cohesive energy is therefore proportional to $\sqrt z$, where $z$
is the atomic coordination number. The
square root form is used assuming that the band energy is the sum of occupied one-electron
levels, and according to the tight-binding model the cohesive energy is
proportional to the
hopping integrals between the d-orbitals. The repulsive potential is summed
in the usual way to describe the repulsive core-core interactions.

Our molecular dynamics (MD) simulations with the {FS} potential
were performed using the Parinello-Rahman~\cite{Parinello} method
and the Nose-Hoover thermostat.\cite{Nose,Hoover} This ensemble
is identified as an isothermal-isotensional ensemble ({NtT}),
\cite{Ray} which allows simulation of fluctuations in the shape
and volume of the sample (here, $N$ is the number of atoms,
 $T$ the temperature, $V$ the volume and $t$ is the sample
 tension).
The  shape and volume thus obtained were used for calculation of
the shear modulus in a canonical ({NVT}) ensemble.  The shear
elastic moduli were calculated using the fluctuations of the
stress tensor. \cite{Rahman}

  The samples used for the simulations contained 2000 atoms,
initially arranged as a perfect bcc crystal of size 10x10x20 unit cells.
Periodic boundary conditions were applied in all three directions.
Point defects were introduced either by the insertion of extra atoms between
the lattice sites (self-interstitials)
or by the removal of atoms from the lattice (vacancies).
Newton's equations of motion were solved using a predictor-corrector algorithm.
\cite{Rapaport, Allen}
Throughout this study, interactive visualization with the AViz
program~\cite{Adler} was implemented.

\section{\label{sec:level1}  Validation of the potential and order parameter }
To learn about the capatibility of the potential, we examined some
physical properties of a  perfect crystal. First, we calculated
the thermal expansion at zero external pressure. We found the
thermal expansion coefficient at low temperatures to be
$\alpha_{c}=(18 \pm 6)\times10^{-6}~K^{-1}$, in reasonable
agreement with the experimental value measured at room temperature
$\alpha_{\exp}=8.6 \times 10^{-6}~K^{-1}$. Next, the thermodynamic
melting temperature for our potential was calculated, using the
method of Lutshko et al.~\cite{Lutsko}, to be $T_m=2220 \pm 10 $ K.
This value is close to the experimental value $T_m=2183$ K,
despite the fact that the FS potential was constructed by fitting
its parameters to room temperature values of various physical
properties of vanadium (lattice constant, cohesion energy, shear
elastic moduli, vacancy formation energy, etc).

\begin{figure}[h]
\centerline{\epsfxsize=10.0cm \epsfbox{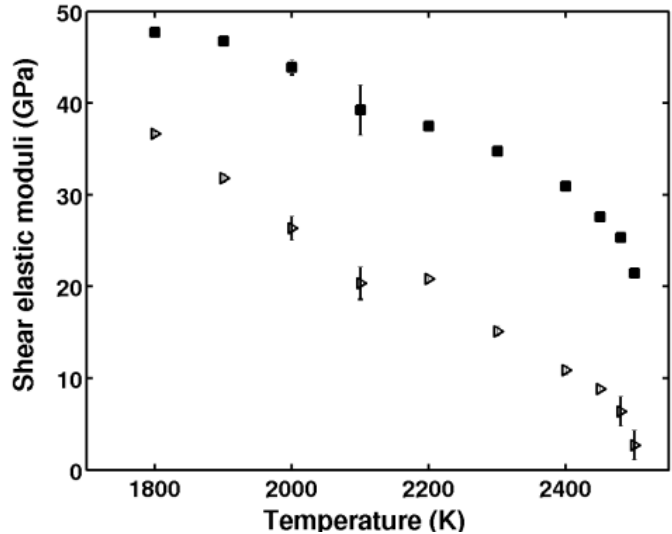}}
\caption{\label{elastic} Variation of $C'$ (triangles) and $C_{44}$ (squares) with temperature.
 The error bars represent statistical uncertainty. }
\end{figure}

In order to test the algorithm we calculated the shear moduli
as a function of temperature.
The shear elastic coefficients decrease with temperature as
shown in Fig.~\ref{elastic}.
 The accuracy of the simulations  was estimated
by monitoring the convergence of the shear elastic moduli calculated along
symmetrically equivalent directions. We found the difference to be approximately
10\%.

Following the validation that our potential can indeed reproduce
the physical properties of a perfect crystal with acceptable
accuracy, point defects were introduced. These point defects are
distributed homogeneously throughout the bulk of the solid. Only
one type of point defects, e.g. vacancies or self-interstitials
were used in each run to avoid their mutual annihilation.

\begin{figure}[h]
\centerline{\epsfxsize=8.0cm \epsfbox{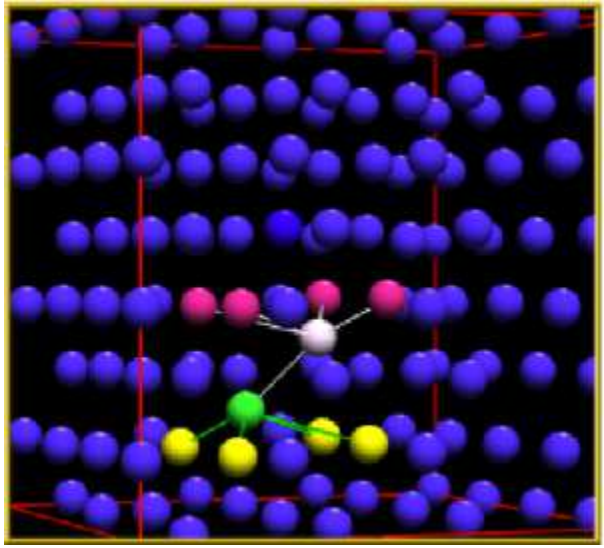}}
\caption{\label{dumb} The most energetically
favored configuration is found to be the
$<011>$  dumb-bell split - interstitial }
\end{figure}

The configurations of atoms in the vicinity of a point defect
inside the bulk at low temperatures was investigated by means of the
simulated tempering method. \cite{Parisi, Yukito} The most energetically
favored configuration of an interstitial was found to be the
$<011>$ {\it dumb-bell split - interstitial} (See Fig. ~\ref{dumb})
with a formation energy of $E_f=4.18  \pm 0.02~eV$.
This formation energy is in agreement with that of previous simulations \cite{Ackland}.

To investigate the temperature dependence of the crystalline order,
we define the structure order parameter $\eta$:
\begin{equation}
\eta=\left < \frac{1}{N^2} \left | \sum_{i=1}^N \cos(\vec k \vec r_i) \right
|^2 +
 \frac{1}{N^2}\left | \sum_{i=1}^N \sin(\vec k \vec r_i) \right |^2  \right >
\end{equation}
where $\vec k=\{0,0,{2 \pi}/{a}\}$ is a vector of the reciprocal lattice,
        $\vec r_i$ is the position of atom $i$,
    $N$ is the number of the atoms in the sample, and
    the angular brackets stands for ensemble average.
For an ideal-crystal lattice at zero temperature, $\eta$ equals unity,
while in the liquid state,  $\eta$  fluctuates near zero.
\begin{figure}[h]
\centerline{\epsfxsize=10.0cm \epsfbox{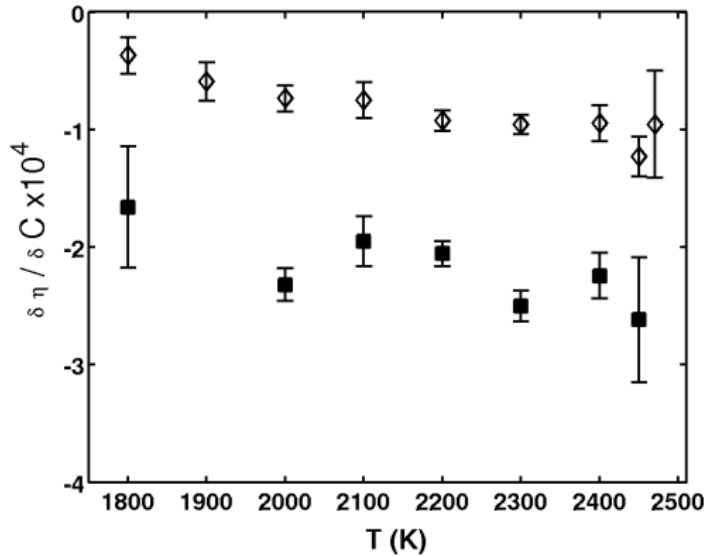}}
\caption{\label{ordd} Influence of point defects: vacancies (diamonds) and
self-interstitials (squares) on the structure
order parameter $ \eta $ as function of temperature. The
concentration of point defects, $C$, is given as a percent of the total
number of atoms. The error bars represent statistical uncertainty. }
\end{figure}

We calculated $\delta \eta / \delta C $, the change of the order
parameter upon the introduction of point defects. Here, $C$ is the
concentration of point defects, given in {\%} of the number of
atoms. Fig.~\ref{ordd} shows the result of this calculation for
small $C$ ,and at different temperatures. The introduction of self
- interstitials results in noticeable decrease of the structure
order parameter (from $\eta \sim 0.6$ to $\eta \sim 0.4$), while
the influence of vacancies is relatively weaker. With increasing
temperature, the order parameter becomes increasingly sensitive to
the introduction of point defects, as evidenced by the increase of
the absolute value of $|\delta \eta /\delta C|$ with temperature.
We believe that this increased sensitivity results from the
increase of the amplitude of thermal vibration of the atoms in the
immediate vicinity of the point defect.
\begin{figure}[h]
\centerline{\epsfxsize=10.0cm \epsfbox{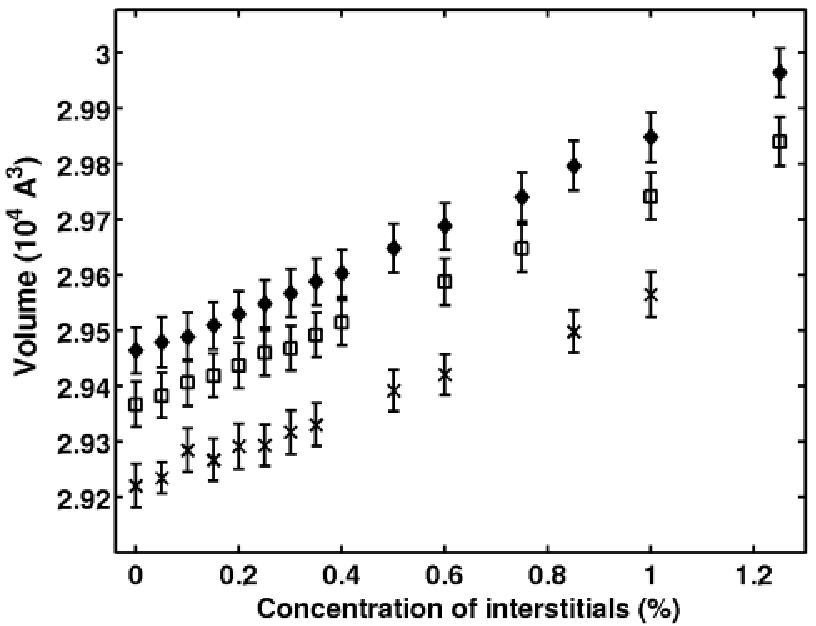}}
\caption{\label{si_vol} Total volume of the sample as a function
of concentration of interstitials at several temperatures: $T=2300~K$ (diamonds),
$T=2200~K$ (squares) and $T=2000~K$ (crosses). The
concentration of point defects is given as a percent of the total number
of atoms. The error bars represent statistical uncertainty. }
\end{figure}
\begin{figure}[h]
\centerline{\epsfxsize=10.0cm \epsfbox{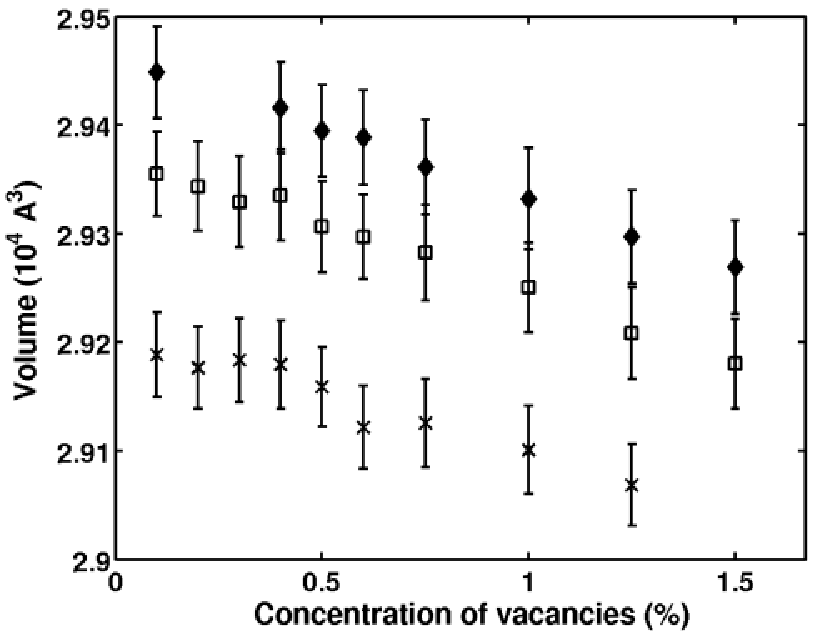}}
\caption{\label{vac_vol} Total volume of the sample as a function
of concentration of vacancies at several temperatures: $T=2300~K$ (diamonds),
$T=2200~K$ (squares) and $T=2000~K$ (crosses). The
concentration of point defects is given in (\%)of the total number
of atoms. The error bars represent statistical uncertainty. }
\end{figure}

\begin{figure}
\centerline{\epsfxsize=10.0cm \epsfbox{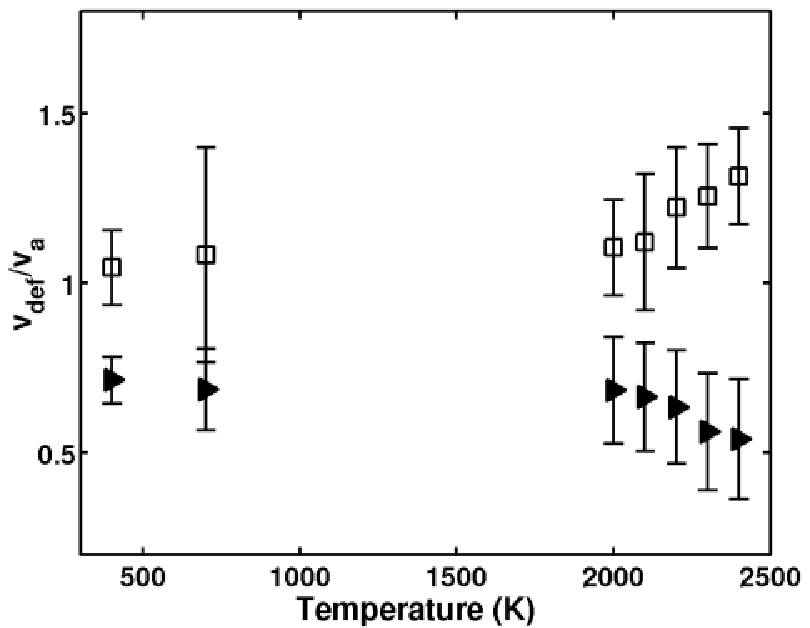}}
\caption{\label{def_vol} The ratio of specific volume of  point
defects to the specific volume of an atom as a function of
temperature: self-interstitials (squares) and vacancies (triangles).
The error bars represent statistical uncertainty. }
\end{figure}

Introduction of self - interstitials increases the volume of the
sample (See  Fig.~\ref{si_vol})
while vacancies decrease the volume, as shown in Fig.~\ref{vac_vol}.
The specific volume of point defects at various temperatures was estimated
using the linear dependence of the volume
on the number of defects, apparent in Figs.~\ref{si_vol} and \ref{vac_vol}.
The volume of the sample at a small number of vacancies can thus be written as
\begin{equation}
 V=(N-N_{va})v+N_{va}v_{va}
 \end{equation}
here, $N$ is the number of atoms in the sample, $N_{va}$ is the
number of vacancies, $v$ is the volume per atom in a perfect
crystal of vanadium and $v_{va}$ is the volume per vacancy. A
similar relation for self-interstitials can be written as
\begin{equation}
V=Nv+N_{si}v_{si}
\end{equation}
where, $N_{si}$ is the number of self-interstitials and  $v_{si}$ is the
volume per interstitial. It is interesting to point out that the
linear dependence appears to hold even at temperatures close to
$T_s$. This may indicate that the concept of a point defect
remains meaningful even under these conditions. The specific
volume of a point defect (in atomic volume units) is shown as a
function of temperature in Fig.~\ref{def_vol}. It is seen that at
temperatures above 2000K these specific volumes change rapidly. To
a large degree, this increase can be accounted for by the rapid decrease of
the elastic coefficients of the crystal in this temperature range.

\section{\label{sec:level1} The bulk melting transition }

The prime goal of our simulations is the investigation of the role
of point defects in mechanical melting. In the simulations of
mechanical melting of fcc metals~\cite{Wolf,Wang1,Kanigel} it was
established that the key parameter controlling melting is the
volume of the crystal. When  the Born criterion is applied to a
superheated crystal lattice it establishes the existence of a
critical volume above which the crystal becomes
mechanically unstable and therefore undergoes a phase
transformation to the liquid state or some other crystal
structure. The critical volume is coupled with a maximum
superheating temperature, $T_s$. Simulations with fcc
metals\cite{Wolf,Wang1,Kanigel} showed that this critical volume,
$v_s$, can be attained by expansion caused either by heating the
crystal, or by doping it with point defects at a constant
temperature which expand the crystal,\cite{Kanigel} or by pure
mechanical dilatation at zero temperature.\cite{Wolf,Wang1} In
this sense the mechanical melting process appears to be universal,
i.e. it is determined only by the sample expansion up to the
critical volume.

In order to verify whether the same scenario holds in the case of
a bcc metal we carried out simulations using samples with  various
concentrations of self-interstitials or alternatively, vacancies.
\begin{figure}
\centerline{\epsfxsize=10.0cm \epsfbox{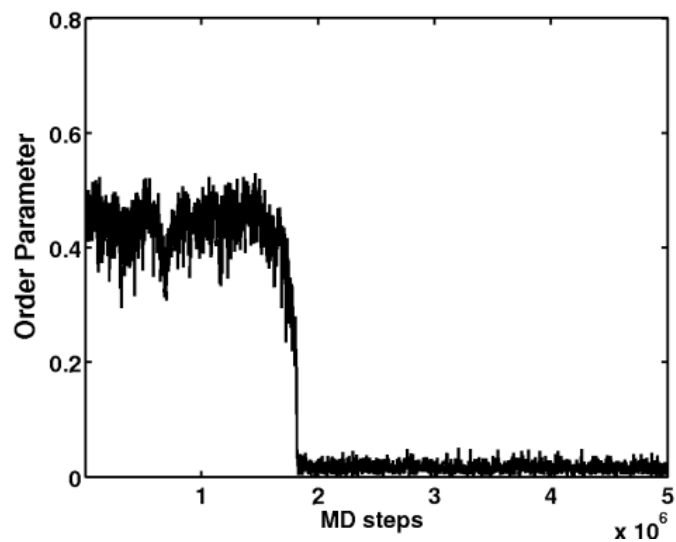}}
\caption{ \label{order_jump} Typical time dependence of the order
parameter during mechanical melting. This particular sample
contained 0.25\% interstitials at temperature $T = 2475 K$. }
\end{figure}
\begin{figure} [h]
\centerline{\epsfxsize=10.0cm \epsfbox{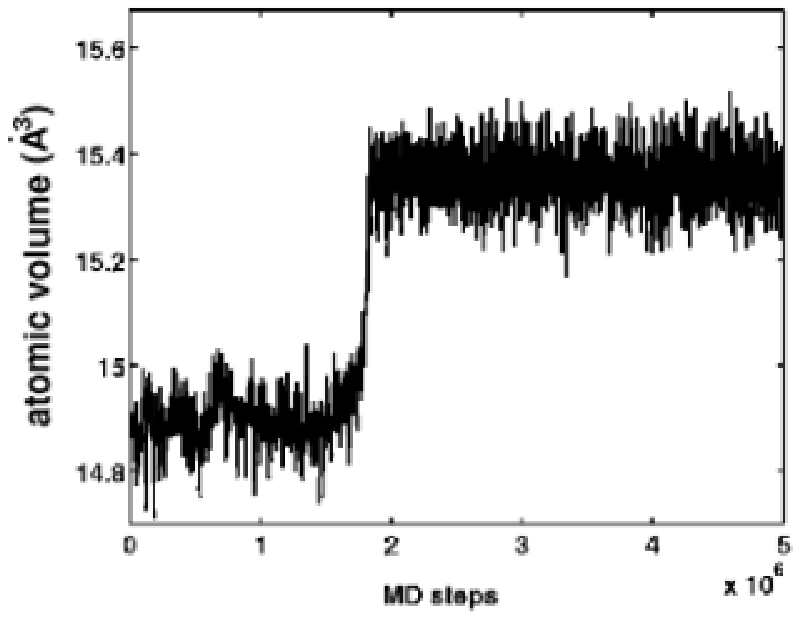}}
\caption{\label{volume_jump} Typical jump of the sample volume
during the mechanical melting transition. This particular sample
contained 0.25\% interstitials at temperature $T = 2475 K$. }
\end{figure}
The initial temperature of each sample was chosen far below the
melting point of a perfect sample, $T \simeq 0.7~T_{s}$. As the
samples were heated by gradually increasing the
temperature, at some point we observed an abrupt decrease of the
structure order parameter (see Fig.~\ref{order_jump}), together with a
simultaneous increase of the total energy and  volume (see  Fig.~
\ref{volume_jump}). This event determines the mechanical melting
temperature. The melting temperature of a sample without
point defects is found to be $T_s=2500 \pm 20$ K. Since MD
simulations are plagued by statistical fluctuations in the
temperature and volume, in practice it is very difficult to
reach the maximum superheating temperature, $T_s$. Therefore, the
accuracy in the determination of $T_s$ in this way is about $\sim
1\% $.

The same temperature, $T_s=2500 \pm 12 $ K, was also found from a
least-squares fit to the temperature dependence of $C'$ as shown
in Fig.~\ref{elastic} It is the temperature where $C'$ goes to
zero. This indicates that as is the case for fcc metals,
homogeneous melting of the bcc metal results from a shear elastic
instability. This particular value of $T_s$ applies to a crystal
of vanadium containing no defects, and is about $280~K$ higher
than the thermodynamic melting point $T_m=2220 \pm 15$ K obtained
for our model using the method proposed by J. F. Luthsko et
al.~\cite{Lutsko}

\begin{figure}
\centerline{\epsfxsize=10.0cm \epsfbox{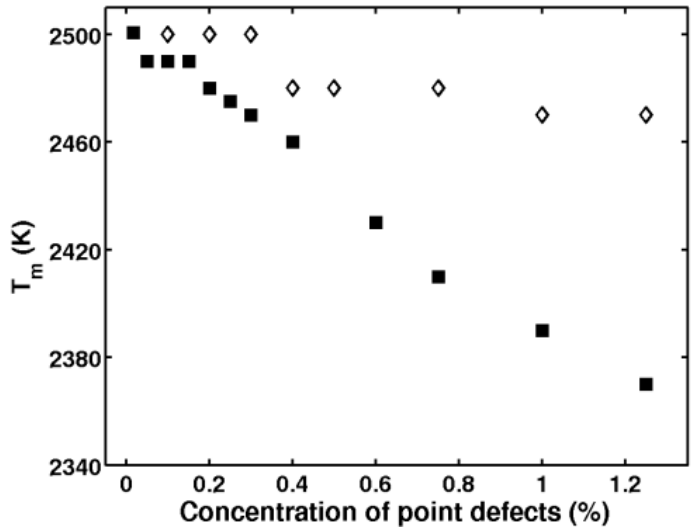}}
\caption { \label{diagram} The influence of interstitials (squares) and vacancies (diamonds)
 on the melting temperature of vanadium under periodic boundary conditions.}
\end{figure}

\begin{figure}
\centerline{\epsfxsize=10.0cm \epsfbox{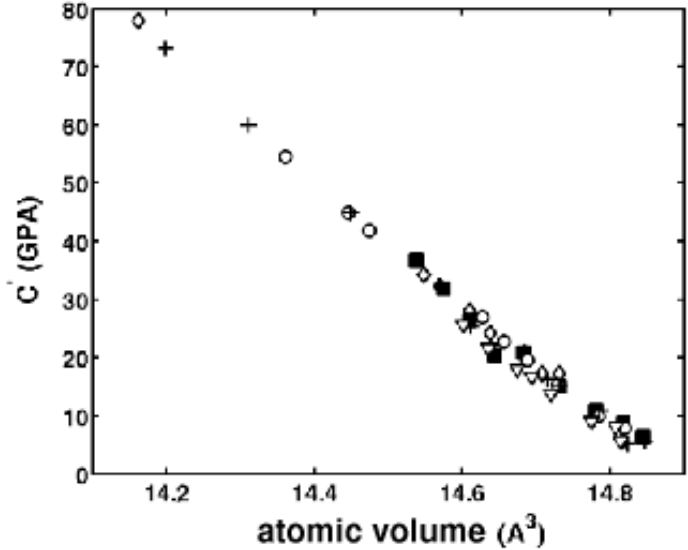}}
\caption{ \label{elast_const_vol}  Plot of the shear modulus $C'$
against specific volume at various concentrations of interstitials:
squares - crystal without impurities (only thermal expansion),
 diamonds - 0.05\% concentration of interstitials,
 circles - 0.1\%, triangles - 0.15\%, crosses - 0.2\%.
 }
\end{figure}

Once point defects are introduced, it is found that $T_s$ becomes
a function of their concentration. Results of simulations
performed at different temperatures and defect concentrations are
summarized in our phase diagram (See Fig.~\ref{diagram}). The fact
that point defects lower the melting temperature has been
confirmed experimentally ( $\gamma$-irradiation lowers
 the melting point of pure metals
by an amount proportional to the dose, and thus to the number of
generated point defects.~\cite{Gorecki, Cahn1}) The lowering of
$T_s$ can be explained as follows: Introduction of self -
interstitials leads to a significant local distortion of the bcc
lattice, and expands the volume of the solid, see
Fig.~\ref{elast_const_vol}. Therefore, a solid containing self -
interstitials reaches its critical volume already at a lower
temperature ( the melting temperature is lower). In contrast, the
effect of vacancies is rather minor, at least if their
concentration is small enough. The same effect of lowering of the
bulk melting temperature induced by interstitials was obtained by
A. Kanigel et al.~\cite{Kanigel} for copper (fcc lattice).
However, at higher concentrations of point defects the decrease of
$T_s$ can not be explained simply by volume expansion. This is
especially notable in the case of vacancies which decrease the
volume, but at high enough concentrations also lower the melting
temperature  (See Fig.~\ref{diagram}). We refer here to the region
in Fig.~\ref{diagram} where the concentration of point defects
approaches 1{\%}. These values are unrealistically large in
comparison with the typical laboratory values $\simeq 0.001\% $.
At these high concentrations, the concept of a single point defect
is unclear and one should perhaps consider clusters, or extended
defects. According to Jin et al. \cite{Jin} extended defects can
act as nucleation centers for melting. Taking this point of view,
the lowering of $T_s$ with defect concentration may result from
the combined effect of (a) volume expansion, and (b) introduction
of nucleation centers for melting. Finally, it should be noted
that the calculated phase diagram is qualitative, because of  the
finite sample size and limited simulation time.

Our results are broadly consistent with  models of defect-induced
melting proposed by Fecht~\cite{Fecht} and Granato.~\cite{Granato}
According to Fecht~\cite{Fecht} melting is driven by the
incorporation of point defects into the lattice. Point defects
increase the probability of heterophase fluctuations of
liquid-like clusters in the defective crystal and lower the Gibbs
energy of the crystalline state. Therefore, the melting
temperature decreases as the concentration of point defects
increases.

The configuration of point defects (self-interstitials)
 in a fcc metals was exploited  by Granato~\cite{Granato}
to construct a model giving the thermodynamic properties of the
crystalline and liquid states in a unified way. He found a large diaelastic softening of the shear modulus
with increasing defect concentration. This leads to a softening of the formation energy of interstitials,
which, together with the large entropy contribution from the interstitialcy resonance modes,
lowers the melting temperature.
 In the above discussion, we have emphasized the role of lattice
instability in establishing a maximum superheating temperature at
zero external pressure. However, due to thermal expansion, any
temperature change is accompanied by a simultaneous change of the
volume. To decouple these two effects, we plot the dependence of
the shear modulus $C'$ on the specific volume in
Fig.~\ref{elast_const_vol}. As this figure shows, the dependence
of $C'$ on the specific volume appears to be universal, in the
sense that the value of $C'$ is the same whether the volume at
which it is calculated was reached at by thermal expansion or by
insertion of point defects. In other words the main effect of
interstitials is to expand the lattice. Using the data plotted in
Fig.~\ref{elast_const_vol} one can extract the value of the
critical volume, $v_s(T_s)$, at which the system melts
homogeneously under the conditions of zero external stress.

Using this method we find $v_{s}=14.87 \pm 0.06 ~\dot A^3$/atom
and the melting temperature $T_s$ for various concentrations of
point defects. The critical volume is close to the specific volume
of liquid vanadium at the thermodynamic melting temperature
$v_{liq}=15.3 \pm 0.05 ~\dot A^3$ and to the experimental value
~\cite{Handbook} of $v_{liq}=15.2 ~\dot A^3$.

Similar results were obtained for copper in MD simulations by J.
Wang et al.~\cite{Wang1,Wang2} and by A. Kanigel et
al.~\cite{Kanigel} It was found that the shear modulus vanishes
at a lattice strain of $a/a_0=1.024$, where $a$ lattice parameter
at  $T_m=1350K$, and $a_0$ is the lattice parameter of copper at
$T_0=300K$. The specific volume ratio of copper is
$(a/a_0)^3=1.07$ which is quite close to the value obtained for
vanadium $v(T_m)/v(T_0)=1.06\pm 0.01$. It is natural to ask
whether the ratio $a/a_0$ is universal, independent of lattice
structure. To answer this question in a definitive manner, it
would be useful to make similar
simulations on other bcc metals.

\section{\label{sec:level1} Summary and conclusions }
In our simulations we observed that each shear elastic modulus is
a continuous and apparently universal function of the specific
volume. The solid lattice can be expanded either by thermal
expansion or by the presence of self-interstitials. The value of
$C'$ at any particular volume is independent of the way by which
this volume was reached, either by thermal expansion alone, or by
any combination of thermal expansion and of expansion due to
interstitials introduced into the sample at a constant
temperature. The elastic energy of the lattice increases until a
critical specific volume $v_{s}$ (close to the specific volume of
the melt) is reached where the shear modulus $C'$ vanishes,
triggering mechanical melting.
 Upon melting, the solid transforms isothermally and discontinuously
(see Figs.~\ref{order_jump} and \ref{volume_jump}).

The process that triggers mechanical melting could be similar to
the one observed by Jin et al. \cite{Jin} in simulations of
the melting of a surface-free Lennard-Jones crystal. There, melting
occurs when the superheated crystal spontaneously generates a
sufficiently large number of extended defects (clusters of
spatially correlated destabilized particles which satisfy the Born
criterion). Those extended defects play the role of surfaces in
initiating the melting. In our simulations, point defects,
especially in large concentration where clusters of defects should
be formed, could act as {\it nucleation centers} for these
extended defects (molten regions) inside the solid.

This paper was devoted to a simulation of the melting process of a
homogeneous bcc metal, and its comparison with a similar process
in fcc metals. It is of great interest to extend these simulations
to heterogeneous melting which involves nucleation of the liquid
phase at some preferred sites of the solid, for example at the
free surface. This study is the subject of a forthcoming paper.

\section{Acknowledgments}
We are grateful to  Dr. G. Wagner, A. Kanigel and J. Tal for
their contribution to this project, and  IUCC for the use of the
computer facilities. This work was supported in part by the Israel
Science Foundation, the German Israel Foundation (GIF), and by
the Technion VPR fund for promotion of research.

\end{document}